\documentclass[preprint,pra,aps,draft,showpacs]{revtex4}

\usepackage{graphicx}% Include figure files
\usepackage{dcolumn}% Align table columns on decimal point
\usepackage{bm}% bold math
\newcommand{\beq}{\begin{equation}}
\newcommand{\eeq}{\end{equation}}
\newcommand{\beqa}{\begin{eqnarray}}
\newcommand{\eeqa}{\end{eqnarray}}
\newcommand{\lam}{\lambda}

\newcommand{\rh}{\rho}

\newcommand{\ka}{\kappa}

\newcommand{\al}{\alpha}

\newcommand{\om}{\omega}

\newcommand{\la}{\langle}
\newcommand{\ra}{\rangle}
\begin{document}

\date{\today}

%\preprint{}

\title{Phonon decoherence of quantum entanglement: Robust and fragile states}

\author{T. Yu}
  %\altaffiliation{}
  %Lines break automatically or can be f%orced with \\
\author{J.\ H.\ Eberly}
\affiliation{Rochester Theory
Center for Optical Science and
Engineering\\
and the Department of Physics \& Astronomy\\
University of Rochester, Rochester, NY 14627, USA\\
% This line
%break forced with \textbackslash\textbackslash
}

\begin{abstract}

We study the robustness and fragility of entanglement of open
quantum systems in some exactly solvable models in which the
decoherence is caused by a pure dephasing process. In particular,
for the toy models presented in this paper, we identify two
different time scales, one is responsible for local dephasing,
while the other is for entanglement decay. For a class of fragile
entangled states defined in this paper, we find that the
entanglement of two qubits, as measured by concurrence, decays
faster asymptotically than the quantum dephasing of an individual
qubit.
\end{abstract}

\pacs{03.65.Yz, 03. 67. -a}

\maketitle

%this is for revtex

\newpage

%%%%%%%%%%%%%%%%%%%%%%%%%%%%%%%%%%%%%%%%%%%%%%%%%%%%%%%%%%%%%%%%%%%%%%%%%%%
\section{Introduction}
%%%%%%%%%%%%%%%%%%%%%%%%%%%%%%%%%%%%%%%%%%%%%%%%%%%%%%%%%%%%%%%%%%%%%%%%%%%
Central to quantum information and quantum computation is the
concept of entanglement of qubits \cite{mc}. In ideal situations,
entangled quantum states would not decohere during processing and
transmission of quantum  information. However, real quantum
systems will inevitably be influenced by surrounding environments.
The interaction between the environment and a qubit system of
interest can lead to decoherence. This is manifest in the loss of
unitary evolution \cite{zurek,zeh}. The decoherence process varies
for different quantum states. On the one hand, the pointer basis
is formed by those states that are unaffected by the environmental
variables, so in this sense they constitute a set of robust states
\cite{zurek}. If such robust states are also entangled, the
entanglement is expected to be stable \cite{dfs1,dfs3,li,
bbk,thew}.

Our intuition strongly suggests that a specified entanglement, as
a nonlocal property of a composed quantum system, should be very
fragile under the influence of the environment. This fragility is
a main obstacle for the realization of practical quantum
computers. Among the various proposals to combat the decoherence
in quantum computing and quantum information processing are ion
traps and nuclear magnetic resonance (NMR)
\cite{cirzol,cirzol1,kie,kie1,viola}.

The main purpose of the present paper is to focus on the issue of
fragility. We show in a specific case that the environment affects
the type of coherence called entanglement quantitatively more
severely than it affects the coherence associated with
off-diagonal matrix elements of a single qubit. We show that this
demonstration can be carried out without approximation for a
non-trivial open system described by a reasonably general
Hamiltonian. Here we write $H_{\rm tot}$ as the sum of
Hamiltonians for the system itself, the environment, and the
coupling between them (we use $\hbar=1$):
\begin{equation}
\label{hamil} H_{\rm tot} = H+ \sum_\lambda\left(g_\lambda
   La^{\dagger}_\lambda + g^*_\lambda L^{\dagger}a_\lambda\right)
   +\sum_{\lambda}\omega_{\lambda} a^{\dagger}_{\lambda}a_{\lambda},
\end{equation}
%\end{widetext}
where $H$ is the Hamiltonian of the system of interest. The
coupling operator $L$ is a system operator coupled to the
environment, and the $g_{\lambda}$ are coupling constants.

We cannot discuss all issues in full generality. Specifically, we
present a toy model in which the environment is represented as a
pure dephasing process. Our final conclusions, based on
calculations of concurrence \cite{woo}, will apply to a two-qubit
pair, and it is fair to say that entanglement of two qubits is a
well-discussed topic \cite{qub}. However, we think that the way in
which entanglement itself decays (or doesn't decay) when a system
is exposed to a nonlocal noisy channel is worth a quantitative
examination.

The organization of the paper is as follows. In Sec. II, we
present an exactly solvable model and the solution of the exact
non-Markovian master equation. In Sec. III and IV, we first
present a two-qubit model and then provide some detailed studies
of robust and fragile entangled states that are initially pure, as
they relax to mixed states. We compare the entanglement decay time
with the local dephasing time in Sec. V. We conclude in Sec VI.
%\newpage

%%%%%%%%%%%%%%%%%%%%%%%%%%%%%%%%%%%%%%%%%%%%%%%%%%%%%%%%%%%%%%%%%%%%%%%%%%%
\section{An exactly solvable model}
%%%%%%%%%%%%%%%%%%%%%%%%%%%%%%%%%%%%%%%%%%%%%%%%%%%%%%%%%%%%%%%%%%%%%%%%%%%%%%

Equation (\ref{hamil}) describes a standard model for open quantum
systems in the system-plus-reservoir framework in which a system
is coupled linearly to a bath of harmonic oscillators, the
excitations of which we can interpret as phonons or photons, for
example. In any event, the bath has distributed eigenfrequencies
$\omega_{\lambda}$ and creation and annihilation operators
$a_{\lambda}^{\dag}, a_{\lam}$ satisfying
$[a_{\lambda},a^{\dag}_{\lambda'}]=\delta_{\lam,\lam'}$. We assume
that the system and the environment are initially uncorrelated:
$\rho(0)=\rho_s(0)\otimes\rho_{bath}(0)$, where $\rho_{bath}(0)$
represents the thermal state of the heat bath at temperature $T$.

In the following, we will consider a specialized model such that
$H= H^\dag$ and $L = L^\dag$ and the two self-adjoint operators
satisfy $[L,H]=i\ka I$, where $\ka$ is a constant, and $I$ is the
identity operator acting on the Hilbert space of the system. A
particular interesting case is when $\ka=0$. That is, $H$ and $L$
commute with each other. Let us note that, except for these
conditions, for the time being, we do not assign operators $H$ and  $L$
any concrete forms.

The exact non-Markovian master equation for the model presented
above takes the following form \cite{yf}: \beqa \label{meq}
\dot\rho_t&=&-i[H,\rho_t] +F(t)[L\rho_t,L]+ F^*(t)[L,\rho_t L]\\
\nonumber
            &&+ G(t)[\rho_t,L]+ G^*(t)[L,\rho_t].
           \eeqa
where
 \beq F(t)=\int^t_0 \alpha(t,s)ds,\ \quad {\rm and}\quad
G(t)=\ka\int^t_0\al(t,s)(t-s)ds,\eeq where
$\alpha(t,s)=\eta(t,s)+i\nu(t,s)$ is the bath correlation function
at temperature $T$:

\beqa
\eta(t,s)&=&\sum_{\lam}|g_\lam|^2\coth\left(\frac{\om_\lambda}{2k_BT}
\right)\cos[\om_\lam(t-s)],\\
\nu(t,s)&=&-\sum_\lam |g_\lam|^2\sin[\om_\lam(t-s)].
\eeqa
Note first that this master equation is in a {\it time-local}
form. So finite memory and thus non-Markovian effects are encoded
in the time-dependent coefficients $F(t)$ and $G(t)$
\cite{dgs,ydgs}. In particular, the $G(t)$ term is a pure
non-Markovian term which does not appear in the Markov
approximation, when $\al(t-s)\rightarrow\delta(t-s)$. This term
gives a non-Markovian phase shift.

Let us now turn to the special case: $\ka=0$, {\it i.e.}, the
system's Hamiltonian $H$ and the coupling operator $L$ commute
with each other. Suppose the states $|n\ra$ are simultaneous
eigenkets of $H$ and $L$:

\beq H|n\ra=E_n|n\ra,\,\,\, {\rm
and}\,\,\,L|n\ra=l_n|n\ra.\eeq
We can explicitly solve the
corresponding non-Markovian master equation (\ref{meq}), and the
solution is surprisingly simple:

\begin{eqnarray}
\label{sol1} \rho_{nm}(t)&=&\la n|\rho_t|m\ra\\ \nonumber &=&{\rm
e}^{-i\left(E_n-E_m\right)t-i(l_n^2-l_m^2)\int^t_0
F_I(s)ds}\\\nonumber &&\times e^{ -(l_n-l_m)^2 \int^t_0
F_R(s)ds}\rho_{nm}(0),
\end{eqnarray}
where the functions $F_R(t)$ and $F_I(t)$ are the real and imaginary
parts of $F(t)$, respectively. The first exponential factor in
(\ref{sol1}) represents a phase shift and the second one
introduces decay, {\it i.e.,} the decoherence effect. In what
follows we always assume that $F(t)$ has an asymptotically
positive real part ensuring that the decoherence is an irreversible
process.

We can see from the solution (\ref{sol1}) that eigenvectors of $L$
are robust states. Precisely, for any initial pure state of the
system $|\psi(0)\ra=|n\ra$, we have
$\rho(0)=|\psi(0)\ra\la\psi(0)|=\rho(t)=|\psi(t)\ra\la\psi(t)|$.

%%%%%%%%%%%%%%%%%%%%%%%%%%%%%%%%%%%%%%%%%%%%%%%%%%%%%%%%%%%%%%%%%%%%%%%%%%%
\section{Two-qubit system}
%%%%%%%%%%%%%%%%%%%%%%%%%%%%%%%%%%%%%%%%%%%%%%%%%%%%%%%%%%%%%%%%%%%%%%%%%%

So far, we have not made any concrete assumptions about the
structure of Hamiltonian $H$ and the coupling operator $L$. In
order to discuss entangled states, we have to specify the
Hamiltonian. This section is devoted to discuss a simple yet
interesting example.

The system we consider consists of two coupled qubits A and B,
where the Hamiltonian for two qubits is taken to be nonlinear and
nonlocal: \beq H=\omega_A\sigma^{A}_z
+\omega_B\sigma^B+J\sigma_z^A\otimes \sigma^{B}_z, \eeq
  and the coupling operator is given by
\beq L=\sigma^A_z + \sigma_z^B\eeq The coupling operator $L$
commutes with the Hamiltonian $H$ and this guarantees that the
energy is conserved at any time. Thus the decoherence  is a pure
dephasing process in which the loss of quantum phase of the system
into the environment is the source of decoherence. The dephasing
is a key issue in practical implementations of quantum computers
\cite{kie,kie1}.

As shown in the last section, the eigenvectors $|n\ra$ of $L$ are
robust states. The solution of (\ref{meq}) for the two-qubit
system is given by

\begin{equation}
\label{sol} \rho_t= \left[
\begin{array}{clcr}
\rho_{11}(0) & \rho_{12}(t) &  \rho_{13}(t) & \rho_{14}(t) \\
\rho_{21}(t) & \rho_{22}(0) & \rho_{23}(0)& \rho_{24}(t) \\
\rho_{31}(t)&\rho_{32}(0)&\rho_{33}(0)&\rho_{34}(t)\\
\rho_{41}(t)&\rho_{42}(t)&\rho_{43}(t)&\rho_{44}(0)
\end{array}
  \right],
\end{equation}
where we have employed the ``standard" eigenbasis:
 $$|1\ra_{AB}=|++\ra, \quad |2\ra_{AB}=|+-\ra, $$
 $$|3\ra_{AB}=|-+\ra, \quad |4\ra_{AB}=|--\ra,$$
and where the matrix elements are:
\beqa \rho_{12}(t) &=&e^{-i(E_1-E_2)t-i4\int^t_0F_I(s)ds}\nonumber \\
&&\times e^{-4\int^t_0F_R(s)ds}\rho_{12}(0), \\
\rho_{13}(t) &=&e^{-i(E_1-E_3)t-i4\int^t_0 F_I(s)ds}\nonumber \\
&&\times e^{-4\int^t_0F_R(s)ds}\rho_{13}(0),\\
   \rho_{14}(t) &=&e^{-i(E_1-E_4)t-16\int^t_0 F_R(s)ds}\rho_{14}(0), \\
\rho_{24}(t) &=&e^{-i(E_2-E_4)t+i4\int^t_0 F_I(s)ds}\nonumber \\
&&\times e^{-4\int^t_0F_R(s)ds}\rho_{24}(0),\\
\rho_{34}(t) &=&e^{-i(E_3-E_4)t+i4\int^t_0 F_I(s)ds} \nonumber \\
&&\times e^{-4\int^t_0F_R(s)ds}\rho_{34}(0).
\eeqa
Here the
eigenvalues of $H$ and $L$ are given by
$$E_1=\om_1+\om_2+J, \quad E_2=\om_1-\om_2-J,$$
$$E_3=-\om_1+\om_2-J, \quad E_4=-\om_1-\om_2+J,$$
and
$$l_1=2, \quad l_2=l_3=0, \quad l_4=-2,$$
respectively.

%%%%%%%%%%%%%%%%%%%%%%%%%%%%%%%%%%%%%%%%%%%%%%%%%%%%%%%%%%%%%%%%%%%%%%%%%%%
\section{Degree of Entanglement:  Robust vs Fragile}
%%%%%%%%%%%%%%%%%%%%%%%%%%%%%%%%%%%%%%%%%%%%%%%%%%%%%%%%%%%%%%%%%%%%%%%%%%

In order to quantify the degree of entanglement, we will adopt the
{\it concurrence } $C$ defined by Wootters \cite{woo}. The
concurrence varies from $C=0$ for an unentangled state to $C=1$
for a maximally entangled state. For qubits, the concurrence may
be calculated explicitly from the density matrix $\rho$ for qubits
A and B: \beq C(\rh)={\rm max}\{0,\lam_1-\lam_2-\lam_3-\lam_4\},
  \eeq
where the quantities $\lam_i$ are the square roots of the
eigenvalues in decreasing order of the matrix \beq
\varrho=\rho(\sigma^A_y\otimes \sigma^B_y)\rho^*(\sigma^A_y\otimes
\sigma^B_y),\eeq where $\rh^*$ denotes the complex conjugation of
$\rh$ in the standard basis.

The most general pure states in the case of the two-qubit model
can be written as

\beq \label{general}|\Psi\ra_{AB}=a_1|1\ra_{AB} +a_2|2\ra_{AB}
  +a_3|3\ra_{AB} +a_4|4\ra_{AB},
 \eeq
 where $\sum_{i=1}^4 |a_i|^2=1$. The
concurrence of the pure state (\ref{general}) is simply given by
\cite{woo} \beq C(|\Psi\ra_{AB})=2|a_2a_3-a_1a_4|.\eeq Thus, the
pure state (\ref{general}) is entangled if and only if \beq
a_1a_4\neq a_2a_3.\eeq

By a robust entangled state we mean one whose entanglement will
not decay to zero in temporal evolution (\ref{sol}). In what
follows, we will prove that the following two special cases with
$a_4=0$ and $a_1=0$:

\beqa \label{rq1}|\psi_1\ra_{AB} &=&a_1
|1\ra_{AB} +a_2 |2\ra_{AB} + a_3 |3\ra_{AB}, \\
|\psi_2\ra_{AB}&=& a_2 |2\ra_{AB} +a_3 |3\ra_{AB} + a_4
|4\ra_{AB}, \label{rqq}\eeqa are robust entangled states if
$a_2a_3\neq 0$. In doing so, let us compute the concurrence of the
density matrix with the initial state (\ref{rq1}). The density
matrix for qubits A and B at time $t$ is given by
\begin{equation}
\label{sta1} \rho_t= \left[
\begin{array}{clcr}
|a_1|^2 &    \rho_{12}(t) &  \rho_{13}(t) & 0 \\
\rho_{21}(t) &  |a_2|^2  & a_2a_3^*  & 0 \\
\rho_{31}(t) &  a_2^*a_3  & |a_3|^2      & 0\\
0 &  0 & 0 & 0
\end{array}
  \right],
\end{equation}
It is easy to check that the concurrence \beq \label{rob}
C(\rh_t)=C(\rh_0)=2|a_2a_3|.\eeq That is, state $|\psi_1\ra_{AB} $
has robust entanglement (which is more than simply being a robust
state). Similarly, we can show, for the entangled pure initial
state (\ref{rqq}), that the degree of entanglement is completely
preserved.

The above result is slightly surprising since the pure states
(\ref{rq1}) and (\ref{rqq}) are the superpositions of the
eigenvectors of $L$ with different eigenvalues. One would expect
that decoherence would eventually degrade the degree of
entanglement. In fact, if we use \beq P(t)={\rm Tr}\rh^2_t\eeq to
quantify the loss of purity of a quantum state, for the initial
pure state (\ref{general}) we have: \beq \label{purity}
P(t)=\sum^4_{i,j=1}|a_i|^2|a_j|^2{\rm
exp}\left[-(l_i-l_j)^2\int^t_0dsF_R(s)\right].\eeq As a measure of
purity of a quantum state, it is easy to see that $0\leq P \leq 1$
where $P=1$ if and only if $\rho$ represents a pure state. From
(\ref{rob}) and (\ref{purity}) we can easily see that in temporal
evolution the purity of the state (\ref{rq1}) deteriorates, but
the degree of entanglement remains constant.

In what follows, we will investigate another class of entangled
pure states whose entanglement tends to vanish under the influence
of the environment. Specifically, we will demonstrate, for the
entangled bipartite pure states (\ref{general}) with $a_3=0$ and
$a_2=0$:
\beqa \label{fra2}|\phi_1\ra_{AB} &=&a_1
|1\ra_{AB} +a_2 |2\ra_{AB} + a_4 |4\ra_{AB},\\
|\phi_2\ra_{AB}&=&a_1 |1\ra_{AB} +a_3|3\ra_{AB} +a_4 |4\ra_{AB}
\label{fra3}, \eeqa that the entanglement will vanish after an
{\it entanglement decay time}  denoted by $\tau_e$. We refer those
states as fragile entangled states.

Now we are in the position to discuss the entanglement decay of
the fragile states by explicitly computing the concurrence. The
density matrix with the initial entangled state (\ref{fra2}) at
$t$ is given by
\begin{equation}
\label{sol2} \rho_t= \left[
\begin{array}{clcr}
|a_1|^2 & \rho_{12}(t) & 0 & \rh_{14}(t) \\
\rho_{21}(t) & |a_2|^2 & 0 & \rh_{24}(t) \\
0 & 0 & 0 & 0 \\
\rh_{41}(t) & \rh_{42}(t) & 0 & |a_4|^2
\end{array}
  \right],
\end{equation}
and the concurrence can easily be obtained \beq C(\rh_t)=
2|\rh_{14}(t)|=2|a_1a_4|{\rm e}^{-16 \int^t_0 ds
F_R(s)}.\label{decay}\eeq The time scale for the entanglement of a
fragile state decaying to zero is determined by the function
$F_R(t)$. For large times, the function $F_R(t)\rightarrow
\Gamma=\int^{\infty}_0\eta(t)dt$. Then in this long time limit,
the entanglement decay time can be identified as \beq
\tau_e^{-1}\equiv 16\Gamma. \eeq

%%%%%%%%%%%%%%%%%%%%%%%%%%%%%%%%%%%%%%%%%%%%%%%%%%%%%%%%%%%
\section{Entanglement decoherence {\it vs} local dephasing}
%%%%%%%%%%%%%%%%%%%%%%%%%%%%%%%%%%%%%%%%%%%%%%%%%%%%%%%%%%%

The dephasing rate of an individual qubit can directly be
estimated from the density matrix for qubit A, which can be
obtained from the density matrix (\ref{sol}) by further tracing
out the variables of qubit B, and vice versa; that is,
$\rh^{A}={\rm Tr}_B\rh,\, \rh^{B}={\rm Tr}_A\rh$. The reduced
density matrix for qubit A is thus obtained from (\ref{sol}):
\begin{equation}
\label{meq1} \rho^A_t= \left[
\begin{array}{clcr}
\rh_{11}+\rh_{22} &    \rho_{13}+\rho_{24}\\
\rho_{31}+\rho_{42} & \rho_{33}+\rho_{44}
\end{array}
  \right],
\end{equation}
Thus, the dephasing rate denoted by $\tau_{\phi}$ for qubit A is
determined by the off-diagonal elements of $\rho^A_t$, \beq
|\rho^A_{12}|=|\rho_{13}+\rho_{24}| \sim {\rm
e}^{-4\int_0^tdsF_R(s)}. \label{dephas} \eeq Similar analysis
applies to qubit B as well. Clearly, the dephasing time depends on
the behavior of the function $F_R(t)$. Similar to the entanglement
decay time, by ignoring the details of the heat bath we can
immediately identify the dephasing time $\tau_{\phi}$ for the
large times as \beq \label{local}\tau_{\phi}^{-1}\equiv
4\Gamma.\eeq

We have thus used the explicit solution of master equation
(\ref{sol}) to evaluate the time development of both the degree of
entanglement and the dephasing rate. As seen from the expressions
(\ref{decay}) and (\ref{dephas}), the entanglement for qubits A
and B and local quantum coherence for an individual qubit, say
qubit A, decay at different rates. We have shown, for a large
class of fragile entangled states, that the entanglement decay
time is shorter than the local dephasing time on which the quantum
coherence of each local qubit is destroyed. Moreover, it can
easily be shown, in the all cases with the entangled initial
states (\ref{general}), that the entanglement decay time is not
longer than the dephasing time, {\it i.e.,} \beq \tau_e
\leq\tau_{\phi}.\eeq

%%%%%%%%%%%%%%%%%%%%%%%%%%%%%%%%%%%%%%%%%%%%%%%%%%%%%%%%%%%%%%%%
\section{Conclusion}
%%%%%%%%%%%%%%%%%%%%%%%%%%%%%%%%%%%%%%%%%%%%%%%%%%%%%%%%%%%%%%%%%
Robust and fragile states have been discussed in some exactly
solvable models. Particularly, for the two-qubit toy model, we
have investigated the dynamics of the robust and fragile entangled
states in terms of a measure of entanglement called concurrence.
We have emphasized and identified that there
exist two different decoherence time scales
--- entanglement decay time and local dephasing time. For the
coupling considered here, we find that the entanglement decay
occurs faster than the local dephasing for a large class of
fragile entangled states. If the size of the active qubits
were to greatly increase, the entanglement decoherence time would be expected to
become exceedingly small, reflecting the classical
limit \cite{comm}. Our example supports one's intuition
about the non-local nature of the entanglement.

Since the amount of entanglement contained in an entangled quantum
state is dependent on the choice of a specific measure of
entanglement, the entanglement decoherence rate is also dependent
on such a choice. We also emphasize that classification of both
robust and fragile  entangled states will depend on the concrete
form of the interaction between qubits and the environment. In
addition, we do not expect that our toy model exhibits all
interesting aspects of the decay processes of the quantum
entanglement. We do believe, however, that the fast decay rate of
quantum entanglement is a generic feature in a variety of physical
processes where decoherence is important.

Finally, it may be worth noting that our two-qubit model allows
two competing processes: creation and annihilation of entanglement
\cite{cra,braun}. Thus, the maximal entanglement generation under
the dephasing process is an interesting problem that will be
addressed in future publications.

%%%%%%%%%%%%%%%%%%%%%%%%%%%%%%%%%%%%%%%%%%%%%%%%%

%%%%%%%%%%%%%%%%%%%%%%%%%%%%%%%%%%%%%%%%%%%%%%%%%
\section*{Acknowledgments}

%\begin{acknowledgments}
We acknowledge grants of financial support from the NSF
(PHY-9415582) and Corning, Inc.
%\end{acknowledgments}
%%%%%%%%%%%%%%%%%%%%%%%%%%%%%%%%%%%%%%%%%%%%%%%%%%%%
%\appendix

%\newpage %Just because of unusual number of tables stacked at end
\bibliography{apssamp}% Produces the bibliography via BibTeX.

\begin{thebibliography}{10}
\bibitem{yf} The details of the derivation of the master equation
(\ref{meq}) will be discussed elsewhere.
\bibitem{comm} For an {\it N}-qubit system, it can be demonstrated
that for the entangled GHZ state $|GHZ\ra=(|0^{\otimes N}\ra +
|1^{\otimes N}\ra)/\sqrt{2}$ the entanglement decay time $\tau_e$
is smaller than the local dephasing time $\tau_{\phi}$ by a factor
$N^2$.
\end{thebibliography}

%\end{references}
\end{document}